\documentstyle[pra,aps]{revtex}
\begin{document}
\preprint{}
\draft
%
%%%%%%%%%%%%%%%%%%%%%%%%%%%%%%%%% TITLE PAGE
%
\title{Dynamics of decoherence in continuous \\
atom-optical quantum nondemolition measurements}
\author{Roberto Onofrio${}^{1*,2\dagger}$ and Lorenza Viola${}^{3,4}$ }
\address{
${}^{1*}$Dipartimento di Fisica ``G. Galilei'', Universit\`a di Padova,
Via Marzolo 8, 35131 Padova, Italy \\
${}^{2\dagger}$Department of Physics and Research Laboratory of Electronics, 
26-259, \\ 
Massachusetts Institute of Technology, 77 Massachusetts Avenue, 
Cambridge, MA 02139\\
${}^3$d'Arbeloff Laboratory for Information Systems and Technology, \\
Department of Mechanical Engineering, 7-040, \\ 
Massachusetts Institute of Technology,
77 Massachusetts Avenue, Cambridge, MA 02139 \\
${}^4$INFM, Sezione di Roma I ``La Sapienza'', Piazzale Aldo Moro 2, 
00185 Roma, Italy 
  }
%\date{\today}
\maketitle
%
%%%%%%%%%%%%%%%%%%%%%%%%%%%%%%%%% ABSTRACT
%
\begin{abstract}
The Lindblad approach to continuous quantum measurements is applied to 
a system composed of a two-level atom interacting with a stationary quantized 
electromagnetic field through a dispersive coupling fulfilling quantum 
nondemolition criteria. 
Two schemes of measurements are examined. 
The first one consists in measuring the atomic electric dipole,  
which indirectly allows one to infer the photon distribution inside the 
cavity. 
The second one schematizes a measurement of photon momentum, which 
permits to describe the atomic level distribution. 
Decoherence of the corresponding reduced density matrices is studied 
in detail for both cases, and its relationship to recent experiments 
is finally discussed. 
\end{abstract}
%
%%%%%%%%%%%%%%%%%%%%%%%%%%%%%%%%% PACS NUMBERS
%
\pacs{03.65.Bz, 42.50.Ar, 42.50.Md}
%
%%%%%%%%%%%%%%%%%%%%%%%%%%%%%%%%% PAPER 
%
%\begin{center}
%To be published in {\sl Physical Review A} {\bf 58} (July issue, 1998)
%\end{center}

\section{Introduction}
Interaction between atoms and photons has been a fundamental issue since 
the early days of  quantum mechanics, and it continues to be a central 
topic expecially in 
connection with controlled manipulation of small numbers of 
photons and atoms at the quantum level \cite{COHEN}. 
In this framework, emphasis has been put recently on repeated measurements 
on atoms and photons in a quantum regime requiring that 
the measurement process be explicitly taken into account. 
It is often desirable that the outcome of a measurement is not 
influenced by the previous ones, which has been found to be possible 
by a clever choice of the measured and measuring systems. 
Indeed, a class of measurements aimed at repeatedly monitoring the same 
observable has been introduced and shown to be compatible with the 
foundations of quantum mechanics, the so-called quantum nondemolition (QND) 
measurements \cite{BRAGINSKY}. 
Many QND schemes have been proposed, and some of them have been implemented, 
to monitor various measurable quantities, {\sl e.g.} 
displacements of macroscopic oscillators below 
the standard quantum limit \cite{CAVES}, photon number in traveling 
\cite{YURKE} or standing \cite{HAROCHE} electromagnetic fields, 
magnetic flux in superconducting interference devices \cite{CALARCO}, 
vibrational energy of an atom confined in a Paul trap \cite{VOGEL}.

From the theoretical point of view, the problem of incorporating quantum 
nondemolition measurements within the language of modern quantum 
measurement theory was not addressed specifically, to our knowledge, 
although important steps were taken in \cite{MILBURN,GAGEN}.
In this article, we analyze a model for QND measurements based on 
Lindblad formalism, specializing it to measurement strategies aimed 
at describing continuous 
quantum nondemolition counting of photons confined in a cavity or 
atoms interrogated to be in a certain internal state. 
The system is described by a density matrix with an evolution 
equation introduced for generic open quantum systems 
\cite{OPEN} and already applied to other, demolitive measurement 
schemes such as the ones involving the quantum Zeno effect in hyperfine 
atomic spectroscopy \cite{PREONTA}, 
optogravitational cavities \cite{ONVI}, superconducting 
circuits \cite{VIONCA}, and trapped ions \cite{VION}. 
In both cases the evolution of the coherence of the monitored system 
is studied, providing a simple picture  
of its decay during continuous QND measurements. 
Even starting from factorized states for atoms and photons, the 
measurement originates an entanglement that is ultimately responsible for 
the indirect decoherence of the observed system through the 
continuous collapse of the state of the probe.
The implications of the model are discussed in connection to recent 
experiments implementing nondemolitive counting of photons in cavity QED 
\cite{HAROCHE} and atoms confined in electromagnetic traps \cite{KETTERLE}.  

\newpage
\section{Quantum nondemolition coupling and open quantum systems}

\begin{center}
{\bf A. General formalism}
\end{center}

In this Section we recall some useful concepts related to quantum 
nondemolition measurements, referring the reader to \cite{CAVES} 
for a more detailed account. 
Although some measurements may occur through direct observation of the same 
observable under study, QND measurements are the most important example of
indirect measurements, whereby the interaction of the system $S$ with another
system $P$ is required.  
The dynamics of the monitored observable $\hat{A}_S$ of the system $S$ is 
inferred through the modifications induced on the actually measured 
observable $\hat{A}_P$ of the probe.
The total Hamiltonian operator is written as 
\begin{equation}
\hat{H}=\hat{H}_S+\hat{H}_{P}+\hat{H}_{int}\;,
\end{equation}
and the evolution of the overall system is described by the density 
matrix equation
\begin{equation}
{d  \over dt}\hat{\rho}(t)=-{i \over \hbar} \left[\hat{H}(t),\hat{\rho}
(t)\right]\;.
\label{RHO}
\end{equation}
The key idea behind a QND measurement is that the subdynamics 
of the observable $\hat{A}_S$ of the monitored system, albeit influencing 
the evolution of the probe, is not affected by this last. 
This nonreciprocity is obtained if the interaction Hamiltonian 
$\hat{H}_{int}$ depends on $\hat{A}_S$ but does not commute with  
$\hat{A}_P$, i.e.
\begin{equation}
[\hat{A}_S, \hat{H}_{int}]=0    \;, \hspace{1cm}
   [\hat{A}_P, \hat{H}_{int}]\neq 0\;. 
\end{equation} 
The subdynamics for the observables $\hat{A}_S$ and $\hat{A}_P$ 
are then ruled, in the Heisenberg picture, by the equations:
\begin{equation}
i\hbar {d \hat{A}_S \over dt}=[\hat{H}_S, \hat{A}_S] \;, \hspace{1cm}
i\hbar {d \hat{A}_P \over dt}=[\hat{H}_P+\hat{H}_{int}, \hat{A}_P]\;.
\end{equation}
In addition, the observable $\hat{A}_S$ should be protected against the 
evolution of all the other system observables whose subdynamics is instead 
unpredictably affected by the measurement, in compliance with the Heisenberg 
principle. 
This requirement restricts the class of the observables that may be monitored
in a QND way, a further sufficient condition being that the 
observable is a constant of the motion in the absence of interaction, 
{\sl i.e.} $[\hat{A}_S,\hat{H}_S]=0$.

This description is still not a measurement model. In particular, the unitary
evolution (\ref{RHO}) is in contrast with the general expectation that a 
measurement process should introduce irreversible signatures into the 
evolution of the system. Actually, the interaction whereby the system $S$ is
monitored by the probe $P$ does not define alone a measurement process. 
As a next step the observable $\hat{A}_P$ has to be registered and, 
in order a measurement is defined univocally, this should correspond to a 
deterministic, classical, amount of information. It is the irreversibility 
implied by this act, with the pointer choosing only one of the many possible 
quantum alternatives, that cannot be accounted for within the closed 
dynamics discussed so far. 
At the beginning of quantum theory, this problem was solved by 
introducing a postulate, which in the von Neumann form 
states the instantaneous collapse of the wavefunction on the eigenstate 
corresponding to the observed eigenvalue \cite{VONNEUMANN}. 
This approach, although applied even in recent times to various situations 
(including the cavity QED case as in \cite{HAROCHE}), has been overcome by a
different description where a unified dynamics for quantum systems is built, 
without additional postulates, using the theory of open quantum systems. 
In this framework, the irreversible nature 
of the measurement is recovered by imagining the act of reading the 
outcome as due to an external environment which continuously interacts with 
the observed system. From a physical viewpoint, any
process in which the interrogations on the observed system are made at a 
repetition rate larger than its intrinsic characteristic frequencies 
may be schematized as a continuous measurement. 
By restricting to so-called {\sl nonselective} ensemble evolutions 
that can be represented in terms of a density operator, 
the most general dynamical law preserving (complete) positivity 
and normalization of the density operator was derived
for a system in interaction with a Markovian environment in the form of a 
so-called Lindblad master equation \cite{OPEN}, later successfully exploited  
in modeling quantum optics experiments \cite{CARMICHAEL}.
Decoherence is introduced into this picture of open system evolutions as the 
dynamical quenching of the off-diagonal density matrix elements, a key 
property first used to explain the absence
of superposition states in a measurement apparatus in \cite{ZUREK}. Within 
this perspective, measured systems are nothing but a special class of open 
quantum systems, the environment being represented by the many-mode field
of the macroscopic measuring apparatus and the measurement process being 
equivalent to repeated instantaneous effect-valued measurements 
\cite{BARCHIELLI,CAVMIL}, or decoherentization kicks given
to the density matrix \CITE{DIOSI}. 
We are thus led to the following master equation for the density of the 
coupled $S+P$ system undergoing a measurement through $\hat{A}_P$:
\begin{equation}
{d \over dt} \hat \rho(t) = 
- {i \over \hbar} \left[ \hat H(t), \hat \rho(t) \right] 
- {\kappa\over2} \left[ \hat{A}_P, [\hat{A}_P, \hat{\rho}(t)] \right]\;. 
\label{RHOK}
\end{equation}
The probe observable, $\hat{A}_P$, plays the role of a Lindblad operator 
representing the influence of the external environment and the parameter  
$\kappa$, with dimensions $[\kappa]=[t^{-1} A^{-2}]$, gives the coupling 
(generally time-dependent) of the probe to the measurement apparatus.
By choosing a continuous function of time $\kappa(t)$ 
a continuous measurement process is obtained. 

\begin{center}
{\bf B. Atom-photon Hamiltonian}
\end{center}

In the following, we will discuss quantum nondemolition counting schemes for 
both photons confined in a cavity and two-level atoms in a given 
eigenstate. By denoting with $\hat{a}, \hat{a}^\dagger$ the standard 
annihilation and creation bosonic
operators, the Hamiltonian of the free single-mode electromagnetic field is 
written as 
\begin{equation}
\hat{H}_{photon}=\hbar \omega \hat{a}^{\dagger} \hat{a} \;, 
\label{HAM}
\end{equation}
while the two-level Hamiltonian can be expressed as
\begin{equation}
\hat{H}_{atom}=\hbar \omega_{ef} \hat{\sigma}_z
\end{equation}
with respect to the basis of the energy eigenstates $|e\rangle, |f\rangle$, 
$\omega_{ef}=(E_e-E_f)/\hbar$. It is convenient to shift the zero of energy 
in order to have $E_f=0$, which is equivalent to choose the following shifted 
two-level Hamiltonian 
\begin{equation}
\hat{H}_{atom}=\hbar \omega_{ef} \bigg( {\hat{I} \over 2} +\hat{\sigma}_z
\bigg)= \hbar \omega_{ef} \hat{\sigma}_+ \hat{\sigma}_-=\hbar \omega_{ef} 
\hat{\Pi}_e \;,
\label{HAMI} 
\end{equation}
where $\hat{\sigma}_+,\hat{\sigma}_-$ and $\hat{\Pi}_e=
\hat{\sigma}_+ \hat{\sigma}_-$ denote the Pauli displacement operators and 
the projector over the excited state $|e \rangle$ respectively.    
Based on the previous considerations, we choose an  
interaction Hamiltonian which is linear in both the photon number operator 
$ \hat{a}^{\dagger} \hat{a}$ and the atomic projector $\hat{\Pi}_e$,
\begin{equation}
\hat{H}_{int}= 2\hbar \gamma  \hat{a}^{\dagger} \hat{a} \hat{\Pi}_e \;,
\end{equation}
the coefficient $2\gamma$, a measurement angular frequency,
quantifying the strength of the quantum nondemolition coupling. 
The occupation probability of level $|e \rangle$ for an atom in a generic 
state $\hat{\rho}^{(atom)}$ is 
$P_e =\mbox{Tr}\{\hat{\rho}^{(atom)} \hat{\Pi}_e\}$.  
We note that, provided the Hamiltonian (\ref{HAMI})
is reinterpreted as single-particle operator for an ensemble of independent 
atoms and collective effects due to quantum statistics or interatomic forces 
are neglected, the probability $P_e$ is related directly to the average 
number $n_e$ of atoms in level $|e\rangle $ through $n_e= P_e n_T$, $n_T$ 
being the total number of atoms. 
Once chosen as system observable $\hat{A}_S$, both the photon number operator
and the atomic occupation probability automatically satisfy the commutation 
relationship with the respective Hamiltonians (\ref{HAM}), (\ref{HAMI}) 
since, in the absence of interactions, they are conserved. 
The total Hamiltonian to be used in (\ref{RHOK}) is written explicitly as
\begin{equation}
\hat{H}=\hbar \omega \hat{a}^{\dagger} \hat{a}+\hbar \omega_{ef} 
\hat{\sigma}_+ \hat{\sigma}_- + 2 \hbar \gamma \hat{a}^{\dagger} \hat{a} 
\hat{\sigma}_+ \hat{\sigma}_- \;. 
\label{TOTHAM}
\end{equation}
We will be working in the representation of the unperturbed (field + atom) 
eigenstates, expanding the density operator as 
\begin{equation}
\hat{\rho}(t)= \sum_{a,b=e,f} \sum_{n,m=0}^{\infty} \rho_{an,bm}(t) 
|a n\rangle \langle b m| \;, 
\end{equation}
where $\rho_{an, bm}(t)=\langle an| \hat{\rho}(t) |bm \rangle$. It may be 
worth to look, for a moment, at the closed evolution of the density matrix
elements. By projecting Eq. (\ref{RHO}) (or Eq. (\ref{RHOK}) with $\kappa=0$),
we find 
\begin{eqnarray}
\label{UNMEAS}
\dot{\rho}_{fn,fm}&=&-i \omega(n-m) \rho_{fn,fm}, \nonumber \\
\dot{\rho}_{en,em}&=&-i(\omega+2\gamma)(n-m) \rho_{en,em}, \\
\dot{\rho}_{fn,em}&=&-i[\omega(n-m)-\omega_{ef}-2\gamma m]\rho_{fn,em} \;,
\nonumber 
\end{eqnarray}
whose solutions are simply rotations of the initial density matrix elements:
\begin{eqnarray}
\label{SOLUNMEAS}
\rho_{fn,fm}(t)&=&\exp{[-i\omega(n-m)t]}\rho_{fn,fm}(0), \nonumber \\
\rho_{en,em}(t)&=&\exp{[-i(\omega+2\gamma)(n-m)t]}\rho_{en,em}(0), \\
\rho_{fn,em}(t)&=&\exp{\{-i[\omega(n-m)-\omega_{ef}-2\gamma m]t\}}
\rho_{fn,em}(0) \;, \nonumber
\end{eqnarray}
and $\rho_{en,fm}(t)=\rho^\ast_{fm,en}(t)$. These equations imply obviously 
that both the average photon number and the average occupation 
probabilities of level $|e\rangle$  are time-independent, 
consistent with the QND nature of the interaction. 
On the other hand, the photon-atom interaction induces additional 
phase-shifts in the density matrix, which are proportional to the strength of 
the coupling $\gamma$ and depend upon the atomic and the photonic state.
These phase-shifts contain the useful information about the system that needs
to be extracted through the measurement on the probe. In the following two 
Sections, we will examine two complementary measurement procedures based on 
this optoatomic coupling. 

\section{QND electromagnetic field measurements via atomic dipole quadrature}

A first class of QND measurements in optoatomic systems is obtained 
by monitoring the photon field using nonresonant atoms as the probe system. 
This corresponds to
\begin{equation}
\hat{H}_S=\hat{H}_{photon}, ~~ \hat{H}_P=\hat{H}_{atom}, ~~ 
\hat{A}_S=\hat{a}^{\dagger} \hat{a}.
\end{equation}
Accordingly, the Hamiltonian (\ref{TOTHAM}) represents the fact that the 
photons in the cavity induce a state-dependent dynamical Stark effect on 
the atom and a selective phase-shift of the atomic wavefunction. 
Due to the requirement of non-commutativity between the probe observable and 
the interaction Hamiltonian,  we discard  any operator proportional to the 
Pauli matrix $\hat{\sigma}_z$ as atomic probe operator 
$\hat{A}_P$. 
On the other hand, in order to model a measurement which is sensitive 
to the dephasing accumulated between the components $e$ and $f$ of the atom 
interacting with the mode, phase-sensitive observables like the quadrature 
components of the atomic dipole operator $\hat{\sigma}_y=(\hat{\sigma}_+-
\hat{\sigma}_-)/2i$, are natural candidates. Indeed, we observe that for a 
proper atomic superposition state,  
$|\psi \rangle=a |e\rangle+ b |f \rangle, \ \ a=
|a| \exp{(i\phi_a)}, 
\ \ b= |b| \exp{(i\phi_b)}, |a|,|b| \neq 0$,  
the average value 
$\langle \sigma_y \rangle=|a| |b| \sin(\phi_{a}-\phi_{b})$ 
is nonzero whenever a relative phase $(\phi_{a}-
\phi_{b})$ is present. By comparison, 
$\langle \sigma \rangle_x=|a| |b| \cos(\phi_{a}-\phi_{b})$, 
maintaining finite values even if the relative phase vanishes. 
We therefore choose 
\begin{equation}
\hat{A}_P={\hat{\sigma}_+-\hat{\sigma}_- \over 2i}
\end{equation}
as the probe observable and by using Eq.(\ref{RHOK}) we get
the following equations of motion:
\begin{eqnarray}
\label{MEASATOM}
\dot{\rho}_{fn,fm}&=&-i \omega (n-m) \rho_{fn,fm}-{\kappa \over 4} 
(\rho_{fn,fm} -\rho_{en,em}), \nonumber \\
\dot{\rho}_{en,em}&=&-i(\omega+2 \gamma)(n-m)\rho_{en,em} 
-{\kappa \over 4} (\rho_{en,em}-\rho_{fn,fm}),  \\
\dot{\rho}_{fn,em}&=&-i[\omega(n-m)-\omega_{ef}-2 \gamma m] 
\rho_{fn,em}-{\kappa \over 4} (\rho_{fn,em}+\rho_{en,fm}), \nonumber 
\end{eqnarray} 
$\dot{\rho}_{en,fm}=\dot{\rho}_{fm,en}^*$.
We note that the evolutions for atomic diagonal and non-diagonal entries 
are decoupled but, at variance with the situations analyzed in 
\cite{PREONTA,ONVI,VIONCA,VION}, the measurement affects all components.  

By introducing the two families of pseudofrequencies $u$ and $w$ defined as
\begin{equation}
u(n,m)=\sqrt{\gamma^2(n-m)^2-{\kappa^2 \over 16}}, \ \ \ 
w(n,m)=\sqrt{\left[\omega_{ef}+\gamma(n+m)\right]^2-{\kappa^2 \over 16}},
\end{equation}
Eqs. (\ref{MEASATOM}) are exactly solved giving 
\begin{eqnarray}
\label{SOLMEAS}
\rho_{fn,fm}(t)&=&e^{-i(n-m)(\omega+\gamma)t-\kappa t/4}
\left\{ \left[\cos ut+i {(n-m)\gamma} {\sin ut \over u} \right] \rho_{fn,fm}(0)
+{\kappa \over 4} {\sin ut \over u} \rho_{en,em}(0) \right\}, \nonumber
\\
\rho_{en,em}(t)&=&e^{-i(n-m)(\omega+\gamma)t-\kappa t/4}
\left\{\left[\cos ut-i {(n-m)\gamma} {\sin ut \over u} \right] \rho_{en,em}(0)
+{\kappa \over 4} {\sin ut \over u} \rho_{fn,fm}(0) \right\}, 
\\
\rho_{fn,em}(t)&=&e^{-i(n-m)(\omega+\gamma)t-\kappa t/4}
\left\{\left[\cos wt+ i [{\omega_{ef}+(n+m)\gamma}] 
{\sin wt \over w} \right]\rho_{fn,em}(0)  
 - {\kappa \over 4} {\sin wt \over w} \rho_{en,fm}(0) \right\} \;, 
\nonumber 
\end{eqnarray}
where the dependence of frequencies $u, w$ on photon numbers $n, m$ is 
understood. 
As expected, Eqs. (\ref{SOLMEAS}) reduce to (\ref{SOLUNMEAS}) when 
no measurement is performed, $\kappa=0$, and for open-system coupling 
$\kappa$ small compared to $\gamma$, the dynamics of the coupled 
$S+P$ system is just weakly perturbed with respect to the closed case. 
In the opposite regime where $\kappa \gg \gamma$ and a proper measurement 
on the signal mode is performed, pseudofrequencies $u$, $w$ tend to become 
purely imaginary, introducing overdamped oscillations and thereby decoherence. 
The time development of the electromagnetic field under the effect 
of the measurement can be inspected by evaluating the reduced density matrix: 
\begin{eqnarray}
\rho_{nm}^{(field)}(t)&=&
\sum_{a=e,f} \rho_{an,am}(t)= 
e^{-i(n-m)(\omega+\gamma)t-\kappa t/4} 
\bigg\{ \Big[\cos ut+\Big(i(n-m) \gamma + 
{\kappa \over 4}\Big) {\sin ut \over u}\Big] 
\rho_{ff}^{(atom)}(0) \nonumber \\
&+ & \Big[\cos ut+\Big(-i(n-m) \gamma +{\kappa \over 4}\Big) 
{\sin ut \over u}\Big] \rho_{ee}^{(atom)}(0)\bigg\}\rho_{nm}^{(field)}(0), 
\end{eqnarray}
where an initially uncorrelated state $\rho_{an,bm}(0)=
\rho_{ab}^{(atom)}(0)\rho_{nm}^{(field)}(0)$ has been assumed. 
It is immediate to recognize that, for every $\kappa$, field populations 
are unaffected by the measurement, i.e. $\rho_{nn}^{(field)}(t)=
\rho_{nn}^{(field)}(0)$, in agreement with the QND nature of the coupling. 
However, the process of acquiring information on the field photon number 
cannot be realized without a back-action on the conjugate field variable, 
specifically an unavoidable degradation of the field phase distribution. 
A direct visualization of the phase evolution is provided by suitable 
phase-coherence indicators. 
We adopt here the formalism of the so-called Pegg-Barnett phase 
distribution discussed at length in \cite{PEGG}: 
\begin{equation}
\Pi^{PB}(\theta,t)=\lim_{s \to \infty} {1 \over 2\pi} 
\sum_{n,m=0}^s \rho_{nm}^{(field)}(t) e^{-i(n-m)\theta}.
\label{BARNETT}
\end{equation}
We have analyzed in detail the evolution arising when the field is initially 
in a coherent state $|\alpha e^{i \phi} \rangle$ with a mean photon number 
$\alpha^2$ and the probe state is an equal superposition of levels $e$,$f$.
As it will become clear in Section V, these choices will enable us a 
straightforward comparison with the results reported in \cite{HAROCHE}. 
We note, however, that equivalent physical insight would be gained from using 
a different standard quantum-optical distribution, the so-called $Q$-function 
\cite{WERNER}. Starting as a sharply peaked function centered at 
$\theta=\phi$, the 
Pegg-Barnett distribution retains its form when $\gamma=0$, the only 
changes reflecting the rotation of the coherent state in phase-space. 
When $\gamma \neq 0$ but $\kappa=0$ the phase distribution is split into 
two components moving at different velocities with relative 
weights proportional to the coefficients  of the atomic wavefunction.
The evolution is strikingly different in the presence of measurement, 
since the phase distribution is progressively scrambled until it becomes 
completely flat when field coherences are asymptotically destroyed, 
$\rho_{nm}^{(field)}(t) \rightarrow 0, n\neq m$. 
The probability distribution (\ref{BARNETT}) is shown in Fig. 1 
for various instants of time and two different values of $\kappa$ 
corresponding to small and strong coupling with the environment.
In the case of weak coupling (solid line), the splitting of the 
initial, well-defined phase distribution, into two components traveling 
with different velocities is still recognizable.  

\section{QND atomic measurements via photon momentum}

A second class of QND measurements is obtained if, according to the notations 
of Section IIA, we choose
\begin{equation}
\hat{H}_S=\hat{H}_{atom}, ~~\hat{H}_P=\hat{H}_{photon}, ~~\hat{A}_S=
\hat{\Pi}_e.
\end{equation}
This corresponds to monitoring the atomic level via an electromagnetic 
field and is realized ordinarily by means of an absorption process, 
i.e. by sending photons resonantly tuned at an energy-level gap.  
This technique is manifestly demolitive for 
the interrogated atoms. An alternative approach
consists in the detection of the phase-shift originated by the 
atom through a non-resonant interaction with a light beam, the atomic sample 
acting as a dispersive medium with a complex refraction index. 
A nice application of this technique has been recently reported  
in \cite{KETTERLE}, where repeated, nondestructive optical imaging 
of an atomic cloud confined in a magnetic trap has been demonstrated. 
The modification of the refraction index, proportional to the atomic 
optical density, induces a phase-shift of the probe field which is 
manifested in turn as a change in the amplitude of its quadrature component. 
The dispersive phase-shifts may be also detected through amplitude 
modulation by means of classical, standard techniques like dark-ground 
or phase-contrast imaging \cite{KETTERLE}.  
In our model, it is natural to choose the following phase-sensitive 
photon operator,  
\begin{equation}
\hat A_P = {\hat a^{\dagger} - \hat a \over 2i},
\end{equation}
as a probe observable. This choice, in addition to fulfilling the QND 
criteria 
established above, corresponds to a measurement configuration which is the 
dual of the one analyzed in Section IV. By exploiting the master 
equation (\ref{RHOK}) again, and by evaluating the new 
Lindblad commutator, equations of motion of the following form are derived: 
\begin{eqnarray}
\label{MEASPHOT}
\dot{\rho}_{an,bm}&=&\dot{\rho}_{an,bm}^{(\kappa=0)}-{\kappa \over 4} 
\bigg[ (n+m+1)\rho_{an,bm}+\sqrt{n(m+1)} \rho_{an-1,bm+1}+
\sqrt{m(n+1)} \rho_{an+1,bm-1} 
\nonumber \\ 
& & - \sqrt{nm} \rho_{an-1,bm-1}-\sqrt{(n+1)
(m+1)} \rho_{an+1,bm+1}  -{1 \over 2} \bigg(\sqrt{n(n-1)} \rho_{an-2,bm} \\
& & +\sqrt{m(m-1)} \rho_{an,bm-2}
+\sqrt{(n+1)(n+2)} \rho_{an+2,bm}+\sqrt{(m+1)(m+2)} \rho_{an,bm+2}\bigg)
\bigg], \nonumber 
\end{eqnarray}
where $a,b=e,f$ and $n,m \geq 0$, and 
$\dot{\rho}_{an,bm}^{(\kappa=0)}$ 
indicates for brevity the appropriate unmeasured contribution, shown 
explicitly in Eqs. (\ref{UNMEAS}). Since, as before, the probe observable 
$\hat{A}_P$ is diagonal in the system variables, no transitions are induced 
by the measurement process in the system. In this case, the $e$ and $f$ 
components evolve independently, but a quite complicated structure of the 
couplings in  the Fock space of the photons is present in general. 
Eqs. (\ref{MEASPHOT}) have been integrated numerically for 
a field coherent state and an equal-weight  atomic  superposition state. 
The reduced density matrix that is appropriate to study the evolution 
of the signal mode during the measurement is the atomic density
obtained by tracing the total density matrix over the photonic 
degrees of freedom,
\begin{equation}
\rho_{ab}^{(atom)}(t)=\sum_{n=0}^{\infty} \rho_{an,bn}(t).
\end{equation} 
By analogy with the previous case, we expect that although no 
perturbation is introduced on the variable which is QND-monitored, 
{\sl i.e.} $\rho_{ee}^{(atom)}(t)=\rho_{ee}^{(atom)}(0)$, atomic coherence 
is eroded due to back-action. 
In Fig. 2 the time dependence of the coherence indicator 
$|\rho_{ef}(t)|$ is plotted for different values of the coupling 
parameter $\kappa$. The evolution of atomic coherence in the closed case 
$\kappa=0$ can be evaluated explicitly for a coherent probe state since, 
from Eqs. (\ref{UNMEAS}), 
\begin{equation}
\label{NONTRIV}
\rho_{ef}(t)={1 \over 2} \,e^{i\omega_{ef}t} \sum_{n=0}^{\infty} 
e^{2i\gamma n t}  \rho_{nn}^{(field)}(0)=
{1 \over 2} \, e^{\alpha^2(\cos{2 \gamma t}-1)} 
\, e^{i(\omega_{ef}t+\alpha^2 \sin 2 \gamma t)}.
\end{equation}
This shows that the quantity ${|\rho_{ef}(t)|}^2$ oscillates 
with angular frequency $2\gamma$ 
between a maximum value equal to $1/4$ and a strictly positive minimum 
equal to $e^{-4\alpha^2}/4$, leading to an  example of nontrivial 
reversible dynamics of the modulus of the atomic coherence. 
In the presence of measurement, atomic coherence damps exponentially to 
zero with a rate proportional to the parameter $\kappa$, still 
preserving the oscillatory behaviour visible in Fig. 2. 

\section{Experimental comparison}

The model developed here, describing decoherence induced by  a continuous QND 
measurement, is characterized completely by the parameters 
$\gamma$ and $\kappa$. 
In this Section we discuss to what extent these parameters can be related to a 
realistic experimental scheme. 
The example we take is the scheme proposed and 
implemented by a group at the Ecole Normale Sup\'erieure \cite{HAROCHE}, where 
atoms detect in a nondemolitive way the photons stored in a high-Q cavity.
In their scheme, the atom is schematized as a system with three Rydberg 
levels $e$, $f$, and $i$, with levels $i$ and $f$ of the same parity, 
opposite to the one of level $e$. 
The frequency $\omega$ of the cavity mode is detuned from the 
$e \rightarrow i$ transition by an amount $\delta=\omega-|\omega_{ie}|$. 
The detuning is large enough to neglect photon absorption, but small 
in comparison to the angular frequency $\omega_{ie}$. 
The presence of photons in the cavity results in a phase-shift of the 
$e$-state relative to the $f$-state. 
We recall that the dynamical frequency shift induced on an atom in level $e$ 
and atomic dipole $d$ and located at point $r$ in a cavity containing $N$ 
photons is
\begin{equation}
\Delta_e(r,N)={\delta \over 2} \bigg\{ {\bigg[ 1+{4 {E(r)^2 d^2} \over 
{\hbar^2 \delta^2}}N \bigg]}^{1/2} -1 \bigg\} \simeq 
{E(r)^2 d^2 \over {\hbar^2 \delta}}N= 
{\Omega(r)^2 \over \delta} N
\end{equation}
where $E(r)$ is the electric field at $r$ and the last approximation 
holds  if absorption processes are made negligible.
The phase-shift is proportional to the vacuum Rabi angular frequency 
$\Omega(r)=E(r) d/\hbar$.
Noticing that $N$ is the eigenvalue of the $\hat{a}^{\dagger} \hat{a}$ 
operator, we can think of the spatially averaged phase-shift as due to the 
effective Hamiltonian  
\begin{equation}
\hat{H}_{int}=\hbar {\langle\Omega^2(r)\rangle \over \delta} 
\hat{a}^{\dagger} \hat{a} \hat{\Pi}_e.
\end{equation}
The parameter $\gamma$ of our model is therefore identified 
as $\gamma=\langle \Omega^2(r)\rangle / 2\delta$. 
Concerning the parameter $\kappa$, the discussion is more elaborated, 
since in the realistic situation analyzed in \cite{HAROCHE} the 
measuring meter is actually a beam of two-level atoms crossing the 
cavity and the effect of the measurement is taken into account using 
the von Neumann collapse for each atomic interaction.
Distinction is made between two different configurations of the probe 
system, corresponding to either atoms with a known initial velocity 
that are registered by the field ionization counters, or atoms which 
are not in a monokinetic velocity state and are not read. 
In the language of quantum measurement theory, this difference can be 
restated in terms of {\sl selective} and {\sl nonselective} measurements 
\cite{PREONTA}, nonselective dynamics being obtained by averaging over all 
the possible states of the probe, in this case over the  velocity 
distribution measurements and the two possible outcomes of the internal level. 
To make explicit comparison with \cite{HAROCHE}, it has been shown there 
that the selective evolution of the field density matrix at the 
$(k+1)^{th}$ atomic detection event  is given by:
\begin{equation}
\rho^{(k+1)}_{nm}(a,v)={{b_a(n,v)b_a^*(m,v)}\over {\sum_n |b_a(n,v)|^2}} 
\rho^{(k)}_{nm}(a,v),
\label{SEL}
\end{equation} 
where $b_a(n,v)$ denotes the amplitude of the component $a$ ($a=e,f$) 
corresponding to a generic atomic velocity $v$ and a photon number $n$. 
As first step towards nonselective measurements, we need to evaluate the 
weighted average over the possible final outcome of the probe, leading to:
\begin{equation}
\rho^{(k+1)}_{nm}(v)=\sum_{a=e,f} {b_a(n,v)b_a^*(m,v)} \rho^{(k)}_{nm}(v)=
\left\{ \sin^2 {\pi v_0 \over 4v}+\cos^2 {\pi v_0 \over 4v} 
e^{-i \epsilon(n-m){v_0 \over v}} \right\} \rho^{(k)}_{nm}(v),
\label{SEMISEL}
\end{equation} 
where the parameter $\epsilon$ measures the accumulated phase-shift per 
photon and  $v_0$ is the atomic velocity corresponding to a $\pi/2$ pulse 
in the Ramsey zone used for interferometric detection of the dephasing.
The next step is a second integration over the atomic velocity 
distribution $P(v)$:
\begin{equation}
\rho^{(k+1)}_{nm}=\int dv \ P(v) \sum_{a=e,f} {b_a(n,v)b_a^*(m,v)} 
\rho^{(k)}_{nm}(v).
\label{NONSEL}
\end{equation}
In Figs. 3  and 4 the Pegg-Barnett phase distribution is plotted 
for a  monokinetic atomic beam, (Eq. (\ref{SEMISEL}))
and a thermal atomic beam, (Eq. (\ref{NONSEL})), respectively.
While in both situations the phase distribution is broadened and 
tends towards 
a flat one, only in the second case the transient does resemble 
the behaviour found for a nonselective measurement 
by using the effective Lindblad approach, Cfr. Fig. 1.
In particular, the velocity of decoherence is proportional to 
the temperature of the atomic beam, i.e. the beam variance. 
This qualitatively agrees with the result established in \cite{PREONTA},  
concerning the direct proportionality 
between the measurement coupling constant and the temperature of the 
bath in which the meter is embedded \cite{PREONTA}. 
Unlike the parameter $\gamma$, it is not possible to infer a simple 
relationship relating $\kappa$ to the various parameters 
of the realistic configuration. 
This shows advantages and disanvantages of the effective Lindblad approach: 
 it is often impossible to relate it completely with the 
experimental set-up; however, a general description of  any measurement 
process is obtained without a detailed knowledge of the actual 
experimental procedure.   
For this reason, the formalism may be easily adapted to the description 
of other relevant schemes, like the single photon-atom coupling in a 
high-finesse Fabry-Perot cavity 
\cite{KIMBLE}, the QND counting of atoms in an optogravitational cavity 
based upon use of evanescent fields \cite{COURTOIS,ASPECT}, and the 
nondestructive imaging of a Bose-Einstein atomic condensate \cite{KETTERLE}.
For the latter system, the dephasing of the atomic coherence induced by the 
measurement process should generate damping of the oscillatory behaviour 
of either a two-species condensate exhibiting Rabi-like oscillations or 
two single-component condensates spatially separated and undergoing coherent, 
Josephson-like oscillations, a situation  similar to the one involving 
a superconducting circuit already analyzed in \cite{VIONCA}.
 
\section{Conclusions}

A model for the description of the decay of coherence in continuous quantum 
nondemolition measurements involving photons and atoms has been developed 
and applied to photon and atom counting. 
Even if initially uncorrelated, atoms and photons become  entangled 
during the measurement and, although conserving their number, they are 
subjected to decoherence due to the back-action via the probe.
The dynamics of the system has been analytically solved in 
the case of photon counting via atomic detection. 
The decay of atomic coherence has been evidenced when the atomic occupation 
probability is measured through the monitoring of the light beam.
Contact has been established with actual experiments in which  
decoherence induced by the measurement process is or could be observable. 
In particular, the nondestructive monitoring of Bose-Einstein 
condensates of atomic 
dilute gas through dispersive imaging, demonstrated in \cite{KETTERLE}, 
will deserve a particular attention, since the influence of the measurement  
on the dynamics of the phase of the wavefunction is a crucial 
issue in the study of macroscopic quantum coherence.
More in general, this model gives constraints on the minimum rate at which 
coherence is progressively destroyed during a continuous measurement process 
even if a quantum nondemolition monitoring is adopted, provided that all 
other possible sources of decoherence have been quenched by proper 
technological improvements. 
As a consequence, similar considerations should be also relevant to 
investigate decoherence dynamics within QED-based quantum computation 
proposals \cite{RAIMOND,KNILL} or quantum control strategies involving 
QND-mediated feedback \cite{TOMBESI}. 
Finally, we point out that in the model described here emphasis has been 
put on considering the whole coupled (system+probe) object as an open 
quantum system in interaction with a measuring environment, the probe 
degrees of freedom being traced out on the total density matrix and 
not dynamically eliminated from the beginning. 
This makes the description formally different from other existing approaches 
\cite{MILBURN,GAGEN}. 
A quantitative discussion of the equivalence between the two strategies 
will be addressed elsewhere \cite{VIOLA}.
%%%%%%%%%%%%%%%%%%%%%%%%%%%%%%%%%ACKNOWLEDGEMENTS
%
\acknowledgments 
We are grateful to Wolfgang Ketterle and Chandra S. 
Raman for a critical reading of the manuscript. 
Use of computing facilities of the INFN, Sezione di Padova, and 
MIT is acknowledged. R.O. is supported by CNR, Italy, through 
the NATO Advanced Fellowships Programme, and MURST, Italy. 
L.V. was supported from University of Padova through the graduate fellowship 
program, and wish to acknowledge the MIT 
Department of Physics for hospitality during the period this work began. 
%
%%%%%%%%%%%%%%%%%%%%%%%%%%%%%%%%% REFERENCES LIST
%

\noindent
*Permanent address; e-mail: onofrio@padova.infn.it

\noindent
$\dagger$Present and corresponding address; e-mail: roberto@amo.mit.edu

\begin{figure}
\caption{Time evolution of the Pegg-Barnett phase distribution for a coherent 
state of the electromagnetic field and various values of the measurement 
coupling constant $\kappa=10^{-1}$  (solid line) and $\kappa=10$ (dashed line) 
in the continuous case. 
The first 20 energy eigenstates have been used as a truncated basis 
for the photon field, leading to a  numerical accuracy of 0.1\%.
The snapshots from above to below are taken at times differing 
by one measurement period defined as $T_m=2 \pi/2\gamma$.}
\end{figure}    

\begin{figure}
\caption{Square modulus of the atomic reduced density matrix 
${|\rho_{ef}|}^2$ versus time (in units of the measurement period 
$T_m$) for various values of $\kappa$. 
The oscillations are present also in the closed case ($\kappa=0$), although 
not visible in the scale, and corresponds to the behaviour of Eq. (25).}
\end{figure}

\begin{figure} 
\caption{Time evolution of the Pegg-Barnett phase distribution for a Von 
Neumann measurement on a coherent state of the electromagnetic field (with 
initial Pegg-Barnett phase distribution as in Fig. 1), corresponding to 
consecutive interrogations with a monokinetic beam at two different 
velocities, $v/v_0=0.3$ (solid) and $v/v_0=0.6$ (dashed). 
The phase is progressively scrambled although its spreading starts from 
well-defined regions.} 
\end{figure}

\begin{figure} 
\caption{Time evolution of the Pegg-Barnett phase distribution for a Von 
Neumann measurement on a coherent state of the electromagnetic field (with 
initial Pegg-Barnett phase distribution as in Fig. 1) with thermal atoms for 
two different temperatures (in arbitrary units) $T=10^{-1}$ (solid) and 
$T=10$ (dashed). 
The scrambling of the phase is faster than in the monokinetic case, 
affecting all the phases, and is proportional to the temperature.} 
\end{figure}
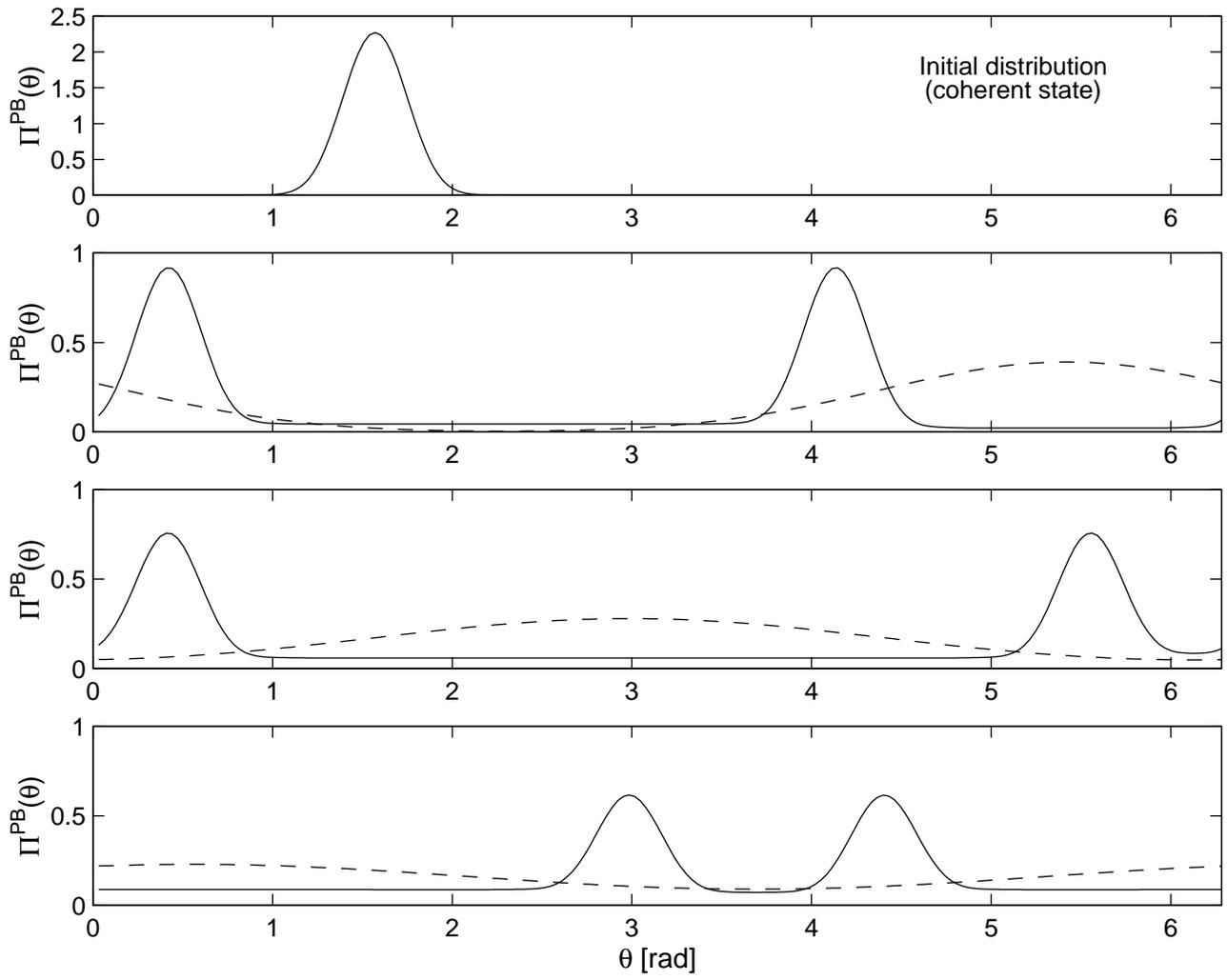
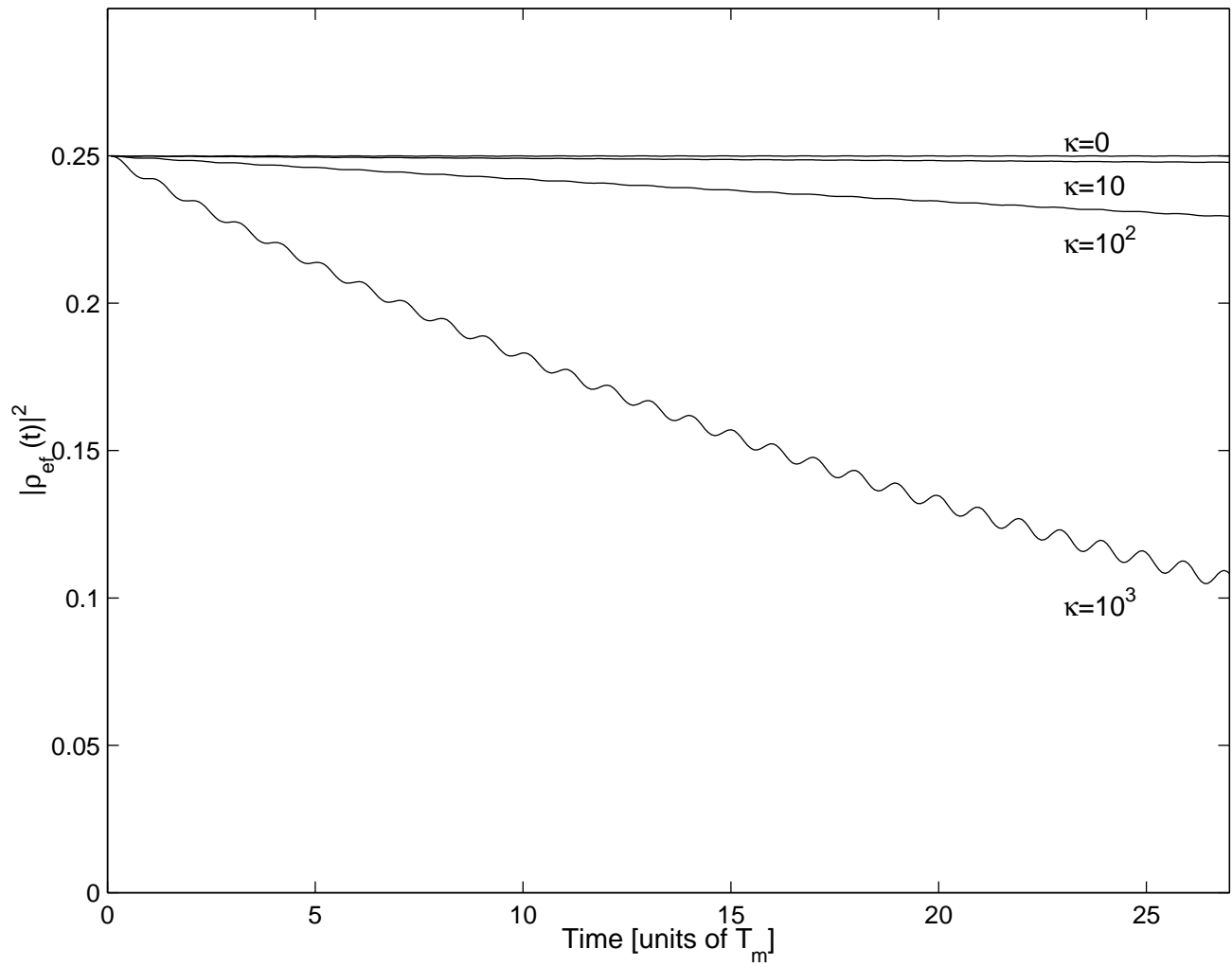
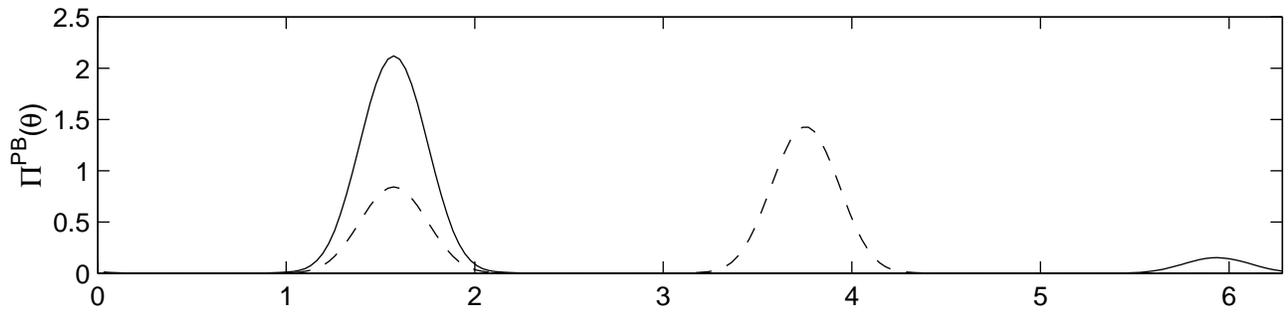
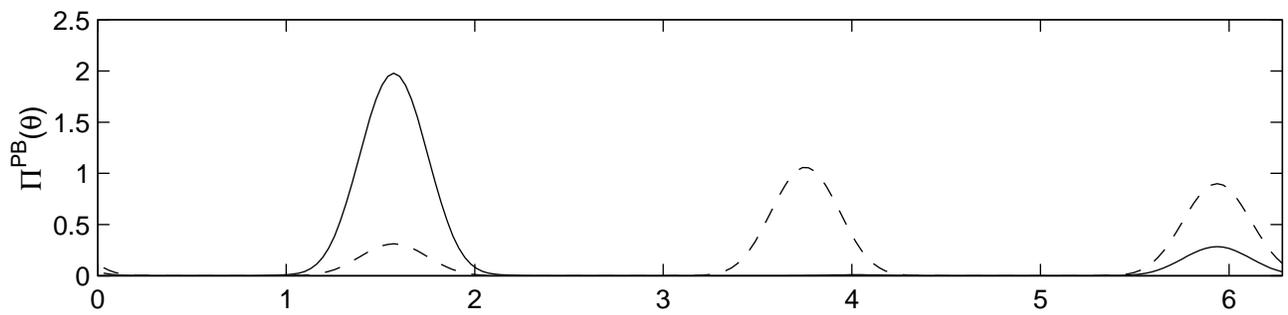
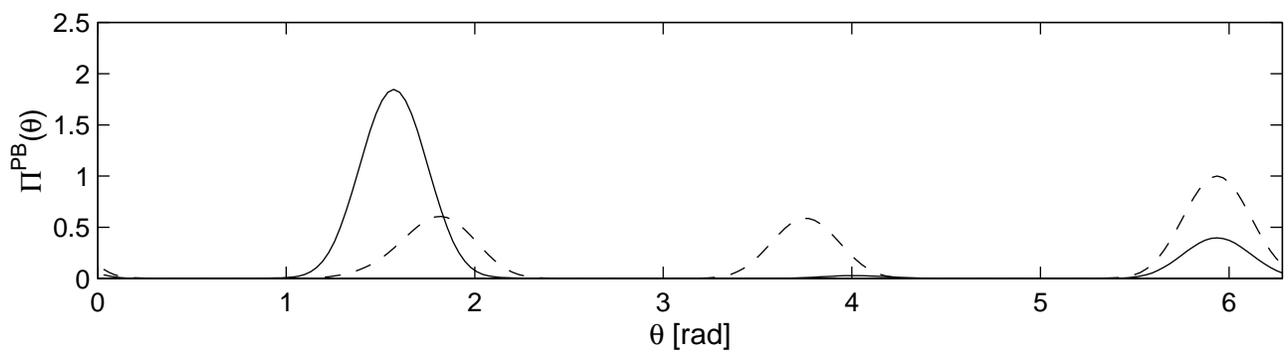
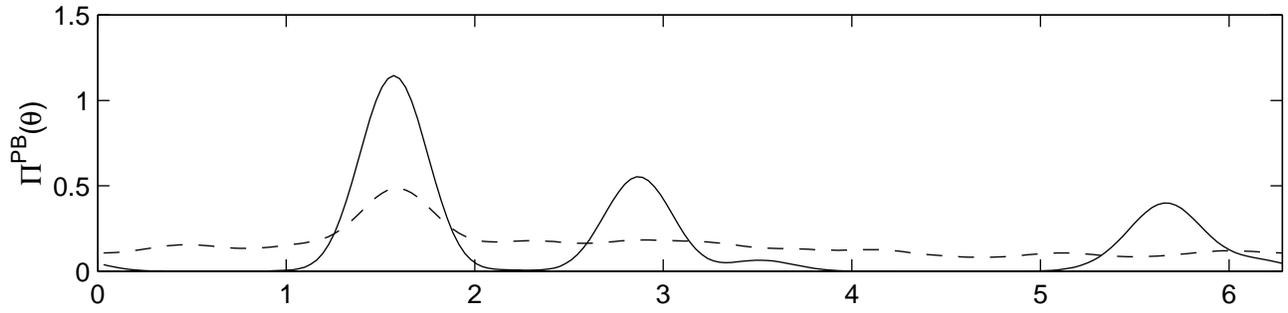
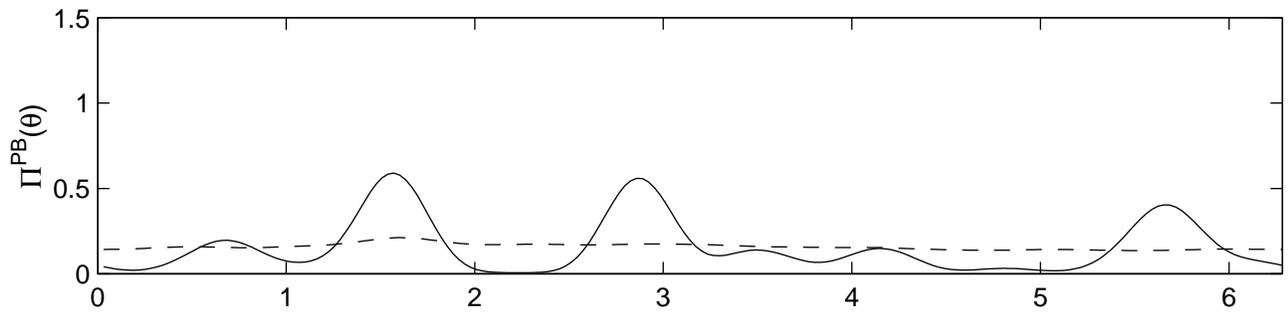
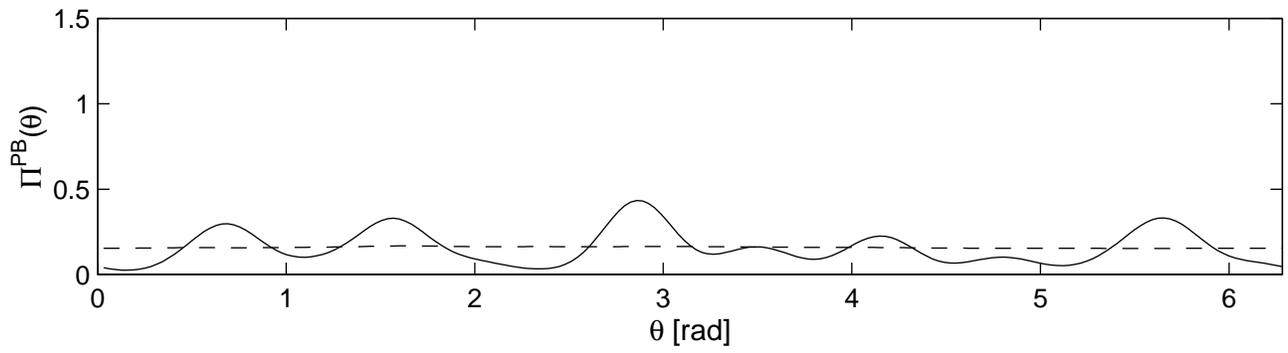
\end{document}